\documentclass[twocolumn,nofootinbib,superscriptaddress]{revtex4-1}
\usepackage{graphicx,epsfig}
\usepackage{amsmath}
\usepackage{amsfonts}
\usepackage{braket}
\usepackage{fancyhdr}
\usepackage{color}
\newcommand{\be}{\begin{eqnarray}}
\newcommand{\ee}{\end{eqnarray}}
\newcommand{\p}{{\cal P}}
\newcommand{\R}{{\cal R}}

\begin{document}

\title{The abundance of primordial black holes depends on the shape of the inflationary power spectrum}
\author{Cristiano Germani}
\email{germani@icc.ub.edu}
\affiliation{Departement de F\'isica Qu\`antica i Astrofisica, Universitat de Barcelona, Mart\'i i Franqu\`es 1, 08028 Barcelona, Spain}
\affiliation{Institut de Ci\`encies del Cosmos, Universitat de Barcelona, Mart\'i i Franqu\`es 1, 08028 Barcelona, Spain}
\author{Ilia Musco}
\email{iliamusco@icc.ub.edu}
\affiliation{Institut de Ci\`encies del Cosmos, Universitat de Barcelona, Mart\'i i Franqu\`es 1, 08028 Barcelona, Spain}
\affiliation{LUTH, UMR 8102 CNRS, Observatoire de Paris, PSL Research University,
Universit\'e Paris Diderot, 92190 Meudon, France}

\begin{abstract}
In this letter, combining peak theory and the numerical analysis of gravitational collapse in the radiation dominated era, 
we show that the abundance of primordial blacks holes, generated by an enhancement in the inflationary power spectrum, 
is extremely dependent on the shape of the peak. Given the amplitude of the power spectrum, we show that the density of 
primordial black holes generated from a narrow peak, is exponentially smaller than in the case of a broad peak. Specifically, 
for a top-hat profile of the power spectrum in Fourier space, we find that for having primordial black holes comprising all of 
the dark matter, one would only need a power spectrum amplitude an order of magnitude smaller than suggested previously 
whereas in the case of a narrow peak, one would instead need a much larger power spectrum amplitude, which in many 
cases would invalidate the perturbative analysis of cosmological perturbations. Finally, we show that, although critical 
collapse gives a broad mass spectrum, the density of primordial black holes formed is dominated by masses roughly equal 
to the cosmological horizon mass measured at horizon crossing.
\end{abstract}

\maketitle


\section{Introduction}
Combination of direct and indirect constraints (for the latest results see \cite{sib,carr, seljak}), indicated that primordial black holes 
(PBHs) could account for all of the dark matter (DM) in the approximate range $[10^{-16},10^{-14}]\cup[10^{-13},10^{-11}]\ M_\odot$.

The observational absence of isocurvature perturbations and non-Gaussianities in the latest cosmic microwave background data 
(the CMB spectrum) favors single field models of inflation {\cite{planck}}. In this context it has been proposed by \cite{garcia1} 
(see also \cite{others,kannike}) that a flattening of the inflationary potential, after the generation of the observed CMB spectra, 
might greatly enhance the power spectrum at scales smaller than those associated with the CMB so as to generate a non-negligible 
abundance of PBHs. 

While PBHs could form by the collapse of statistical fluctuations of curvature perturbations generated during inflation, the usual slow-roll 
approximation, which well describes CMB physics, fails in this case \cite{kannike}, so that a more careful analysis must be performed, 
as discussed recently in \cite{germani,Hu, ballesteros}.  Additionally, the abundance of PBHs depends on the amplitude of the inflationary 
power spectrum and a threshold ${\cal P}_c$. This threshold is related to the minimum amplitude of initial curvature perturbations eventually 
collapsing to form black holes. 

Recently there has been some confusion about the correct estimate of $\p_c$: for example, in \cite{garcia1} and \cite{garcia2} a rather 
small value of $\p_c\sim {\cal O}(10^{-1})$ has been mistakenly equated to the analytical estimate of the critical value $\delta_c$ for the 
integrated density perturbations \cite{harada}. A larger value of $\p_c\sim {\cal O}(1)$ \cite{germani} was obtained by incorrectly converting 
the critical amplitude of the integrated density perturbations into $\p_c$, as in \cite{ballesteros} (in the realm of effective field theories) 
and in \cite{cicoli} (within explicit string theory realizations).

In the present paper, we show using peak theory \cite{peak}, that all previous estimates of $\p_c$ are actually inconsistent with the numerical 
simulations of PBH formation \cite{Niemeyer, Shibata, muscop, harada2, muscoin}, whether or not the PBHs comprise the whole of the DM. 
The key point is that the threshold $\p_c$ is not universal but instead strongly depends on the shape of the inflationary power spectrum. 

Peak theory was already used in \cite{malik} to calculate the abundance of PBHs, without considering the relation between the shape of the 
inflationary power spectrum and the threshold of the energy density peak. In the following we propose an improved procedure for calculating 
the PBH abundance taking into account also the effect of the shape of the power spectrum.


\section{Cosmological perturbations and PBH formation}\label{PBH_formation}
In the radiation dominated era, PBHs could be formed by sufficiently large cosmological perturbations collapsing after re-entering the 
cosmological horizon.  Assuming spherical symmetry, such regions can be described by the following approximate form of the metric at 
super-horizon scales 
\be
ds^2 \simeq -dt^2+a^2(t)e^{2{\cal R}(r)} \left[ dr^2 + r^2d\Omega^2 \right] 
\label{pert_metric}
\ee
where $a(t)$ is the scale factor while $\R(r)$ is the comoving curvature perturbation. In this regime the curvature perturbation is 
non-linearly conserved \cite{Lyth:2004gb} and, from the Einstein equations, in the gradient expansion approximation \cite{muscoin, grad}, 
one has
\be\label{rz}
\frac{\delta\rho}{\rho_b}(r,t) \simeq - \frac{1}{a^2H^2} \frac{8}{9} \, e^{-5\R(r)/2} \,\nabla^2 e^{\R(r)/2}  
\ee 
where $\nabla^2$ is the flat space laplacian, $H\equiv\dot{a}(t)/a(t)$ is the Hubble parameter and $\rho_b(t) = 3 M_p^2 H^2(t)$ is the 
background energy density. 

In the metric \eqref{pert_metric} the areal radius is given by \mbox{$R(r,t) = a(t)r e^{\zeta(r)}$} and the amplitude of a cosmological 
perturbation can then be measured by the mass excess within a spherical region of radius $R$ as
\be
\frac{\delta M}{M_b} \simeq \delta(r,t) \equiv  \frac{1}{V_b} \int_0^R 4\pi\frac{\delta\rho}{\rho_b} R^2 dR \ ,
\ee
where $M_b(r,t) = V_b(r,t)\rho_b(t)$ is the background mass within the spherical volume $V_b(r,t) = 4 \pi R^3(r,t) / 3$. 

As explained in \cite{muscoin}, a PBH can be formed when the maximum of the \emph{compaction function} 
\mbox{$\mathcal{C}\equiv 2G\,\delta M(r,t)/R(r,t)$} is larger than a certain threshold value. This prevents the over-density bouncing 
back into the expanding Universe. At super-horizon scales, when the maximum of $\mathcal{C}$ is located well outside the 
cosmological horizon, this quantity is conserved and is related to the mass excess by
\be 
\delta(r,t) \simeq \left( \frac{1}{aHr} \right)^2 \mathcal{C}(r) \,.
\ee
The location of this maximum, called $r_m$, is an important quantity measuring the characteristic scale of the density perturbation. 
Comparing different profiles in terms of $r/r_m$ one has that similar shapes measured in these units have similar values of the mass 
excess threshold \mbox{$\delta_c\equiv\delta_c(t_m,r_m)$}, where $t_m$ is defined by \mbox{$a(t_m)H(t_m)r_m=1$}. This identifies 
the so called ``horizon crossing'' measured in real space, but one should bear in mind that $\delta_c$ is calculated using the approximation 
of $\delta(r,t)$ at super-horizon scales.

Although the threshold $\delta_c$ characterizes the mass excess needed to form PBHs, it is the critical value of the peak 
$\delta\rho_c(0)/\rho_b$ that plays a crucial role for computing their cosmological abundance as we shall see in the next section. 

By performing a detailed numerical study it has been found that, depending on the initial profile of the energy density,
the threshold $\delta_c$ is in the range {\mbox{$0.41 \lesssim \delta_c \leq 2/3$}}, which is related to the range of critical values of the 
energy density calculated at the center of the over-density, with $\delta\rho_c(0)/\rho_b\geq 2/3$ (see \cite{muscoin} for more details).

 
\section{Applications of peak theory}

\subsection{The average density profile}
In the previous section we have discussed the conditions for which a single perturbation is able to form a PBH. In this section we will 
apply this knowledge to the cosmological perturbations generated during inflation. 

Cosmological perturbations are of quantum origin and therefore their shapes and amplitudes are statistically distributed. In particular 
$\Delta\equiv \delta\rho/\rho_b$ is a statistical variable and since we assume that perturbation theory applies during inflation, the mean 
value of $\Delta$ and thus the gradient of $\R$ and its amplitude, are very small. As discussed earlier however, to form PBHs we do need 
``large'', i.e. non-linear, values of $\Delta$. Therefore, we will need to search for large perturbations (peaks) away from the mean value. 
Assuming that both $\Delta$ and $\R$ are approximately gaussian variables\footnote{ Because the formation of a PBH is a rare event, 
in principle the abundance of PBHs can be modified by non-Gaussian contributions to the statistics of the primordial curvature perturbations \cite{vicente,kehagias2}. Whether or not these non-Gaussianities are important is a model dependent question which is still under debate 
(for more details see \cite{sasaki, vanin, diego,vicente}) and will not be addressed in this paper. }, with the help of peak theory \cite{peak}, 
those  peaks will be described only by the variance of $\Delta$, which is completely dominated by the two-point function of $\R$ via the 
linearised relation
\be\label{approx}
\frac{\delta\rho}{\rho_b}\simeq - \frac{1}{a^2H^2} \frac{4}{9} \, \nabla^2 \R  \ .
\ee  
Higher correlators will then be suppressed by higher powers of the power spectrum\footnote{In a paper which appeared on the same day 
as ours \cite{Jaume}, these corrections were evaluated finding that the variance of $\Delta$ is slightly larger than the one found here.} of $\R$. 
In Fourier space we then have
\be\label{xi}
(2\pi)^3P_{\Delta}(k,t)\delta(k,k')&\equiv&\langle\Delta(k,t)\Delta(k',t)\rangle \simeq 
\left(\frac{k}{aH}\right)^4 \frac{16}{81}\times\cr&\times&(2\pi)^3 \delta(k+k'){\frac{2\pi^2{\cal P}(k)}{k^3}}\ ,
\ee 
where we have used a standard definition of the curvature perturbation power spectrum ${\cal P}(k)$ \cite{mukhanov}. Finally, we can then 
define the moments of $P_{\Delta}(k,t)$ as
\be
\sigma^2_j(t)\equiv\int \frac{k^2 dk}{2\pi^2}P_{\Delta}(k,t)k^{2j} \,.
\ee

The density of PBHs at the moment of formation must be much smaller than the density of the background radiation, otherwise they will 
dominate the present Universe when it becomes matter dominated. For this reason the peaks generating PBHs must be rare and, to a good 
approximation, can be considered spherical. Non-sphericity of the peaks would be obtained by the interaction of different adjacent 
over-densities \cite{peak}.

The observed super-horizon density profile is constructed by using the multivariate Gaussian distribution of the (real space) random 
field $\Delta(r,t)$. Following \cite{peak} the super-horizon averaged density profile is measured in terms of the relative amplitude of the the 
peak defined as \mbox{$\nu\equiv\frac{{\cal F}(0)}{\tilde\sigma_0}\gg1$}, which implies that peaks are rare. Then the mean over-density 
profile per given central value is
\be\label{Fn}
F(r,t) \simeq \frac{{\cal F}( r)}{a^2 H^2}
\ee
with
\be
{\cal F}( r)\equiv{\cal F}(0)\frac{\xi( r, t)}{\xi(0,t)}\ ,
\ee
where \mbox{$\tilde\sigma_0\equiv\sigma_0(t)a^2H^2$} and ${\cal F}(0)/(a H)^{2}$ is the amplitude of the over-density at the center of the 
profile and 
\be\label{xi2}
\xi(r,t)=\frac{1}{2\pi^2\times (2\pi)^3}\int dk  k^2 \frac{\sin\left(k r\right)}{k r}P_{\Delta}(k,t)\ .
\ee
In this limit the number density of peaks corresponding to a given amplitude ${\cal F}(0)$, in the comoving volume, is
\be\label{beta}
{\cal N}_c(\nu)=\frac{k_*^3}{4\pi^2}\nu^3 e^{-\nu^2/2}\theta(\nu-\nu_c)\ ,
\ee
where  $k_*\equiv \frac{\sigma_1}{\sqrt{3}\sigma_0}$ and, at super-horizon scales, $\nu$ is {\it time independent}. The critical value $\nu_c$ 
discriminates between perturbations forming black holes ($\nu>\nu_c$) and perturbations dispersing into the expanding Universe ($\nu<\nu_c$).
\footnote{ The spreading of the profiles can be estimated following section VII of \cite{peak}:\[ \frac{\sqrt{\langle \left(\Delta(r,t_m)-{\cal F}(r)r_m^2\right)^2\rangle}}{{\cal F}(r)r_m^2}\simeq  \frac{1}{\nu}\frac{\sqrt{1-\psi(r)}}{\psi(r)} \] where 
$\psi(r)\equiv\frac{\xi(r,t)}{\xi(0,t)}$. Since $\nu\gg 1$ our approximation of considering only the threshold value of the mean 
profile, instead of the mean threshold, is a good one around the peak. There would be some small effects related to the 
edge of the profile, but since they are small for the calculation of the threshold \cite{muscoin}, we neglect them here (for more 
details we refer to \cite{Jaume} where these effects have been estimated). }

\subsection{Abundance and mass spectrum of PBHs}
The number of ``sufficiently large '' peaks at super-horizon scales gives us the number of PBHs formed once the over-density crosses 
the horizon. Then the number density of PBHs in physical space, at the moment of formation, is given by
\be
{\cal N}_p(\nu)=\frac{{\cal N}_c(\nu)}{a(t_f)^3}\ ,\nonumber
\ee
where $t_f$ is the time when the PBHs are formed. Note that $k_*/a$ is not dependent on the rescaling of the scale factor and so the 
same is also valid for ${\cal N}_p(\nu)$, as it should be. Finally, we are now able to define the density of PBHs of a given mass $M_{PBH}(\nu)$ 
at formation to be
\be
\rho_{PBH}(\nu)\simeq M_{PBH}(\nu){\cal N}_p(\nu)\ .
\ee
The relative density of PBHs that would still exist today, measured at formation with respect to the background energy-density, is 
\be\label{betaf}
\beta_f\equiv \int^{\infty}_{\nu_{min}}\frac{\rho_{PBH}(\nu)}{\rho_b(t_f)}d\nu 
\ee
where $\rho_b(t_f)=3 M_p^2 H^2(t_f)$ and $M_p$ is the Planck mass. The lower limit $\nu_{min}$ corresponds to 
\mbox{$M_{min}\sim 10^{15}\ {\rm g}$} which is the mass of PBHs that would already have evaporated by now. To match the abundance 
of PBHs with the observed DM, one should have \mbox{$\beta_f\simeq 10^{-8}\sqrt{\frac{M_{PBH}}{M_\odot}}$}, as can be seen for example 
in \cite{cicoli}. 

For given $\nu$, the PBH mass is well approximated by the scaling law for critical collapse \cite{Niemeyer,muscop}
\be\label{mpbh}
M_{PBH}\simeq{\cal K} M_H(t_m)\left(\frac{\tilde\sigma_0}{a_m^2 H_m^2}\right)^\gamma \left(\nu-\nu_c\right)^\gamma\ ,
\ee
where for radiation $\gamma\simeq 0.36$, ${\cal K}\sim{\cal O}(1)$ is a numerical coefficient that depends on the specific density profile 
and $M_{H}(t_m)\equiv 4\pi \frac{M_p^2}{H_m}$ is the horizon mass measured at horizon crossing.  

Finally we have 
\be\label{int}
\beta_f&\simeq&\frac{{\cal K}}{3\pi}\left(\frac{k_*}{a_m H_m}\right)^3\left(\frac{\tilde\sigma_0}{a_m^2 H_m^2}\right)^\gamma \nu_c^{3+\gamma}\ {\cal I}(x_{\rm min})\ ,
\ee
where
\be
{\cal I}(x_{\rm min})\equiv\int_{x_{\rm min}}^\infty \frac{a_f}{a_m}x^3 \left(x-1\right)^\gamma e^{-\frac{x^2}{2 \nu_c^{-2}}}dx \ ,\nonumber
\ee
and $x\equiv\frac{\nu}{\nu_c}$. Numerical simulations show that $a_f$ is only weakly dependent on $\nu$ \cite{muscoin}, 
giving approximately $a_f/a_m\simeq3$, and therefore we take this factor out from $\cal I$.

Assuming that the horizon mass at formation is much larger than $10^{15}\ {\rm g}$, since otherwiseno relevant PBH abundance would be generated\footnote{We are envisaging here that these PBHs would account for all the DM, or for a significant part of it.}, in the large $\nu_c$ 
limit, one can approximate the previous integral with its saddle point at $\nu_s\simeq\nu_c+\frac{\gamma}{\nu_c}$ as was done in \cite{carr2}, 
obtaining
\be\label{bf}
\!\!\!\!\!\!\!\!\!\!\!\!\!\!\!\beta_f\simeq \sqrt{\frac{2}{\pi}}{\cal{K}}\left(\frac{k_*}{a_m H_m}\right)^3 
\left(\frac{\tilde\sigma_0}{a_m^2 H_m^2}\right)^\gamma  \nu_c^{1-\gamma}\gamma^{\gamma+1/2}e^{-\frac{\nu_c^2}{2}} \!\!\!\!
\ee 
The error from using this approximation grows slowly with $\nu_c$ but always stays around $10\%$, for $\nu_c={\cal O}(10)$.

If the linear approximation applies (\mbox{$\nu_c\gg 1$}), the density of PBHs would be typically peaked at the saddle point of 
 \eqref{int} which, inserting it into \eqref{mpbh}, gives
\be\label{Mpeak}
M_{PBH}(\nu_s)=4\pi{\cal K} \frac{M_p^2}{H_m}\left(\frac{\tilde\sigma_0}{a_m^2 H_m^2}\right)^\gamma \left(\frac{\gamma}{\nu_c}\right)^\gamma\ .
\ee
Although $M_{PBH}(\nu_s)$ still gives the right order of magnitude for the black hole masses dominating the DM density, the square 
root of the variance of (11) is numerically calculated to be about $1.2\, M_{PBH}(\nu_s)$.

\begin{figure*}[t!]
\centering
\vspace{-1.cm}
\includegraphics[scale=0.40]{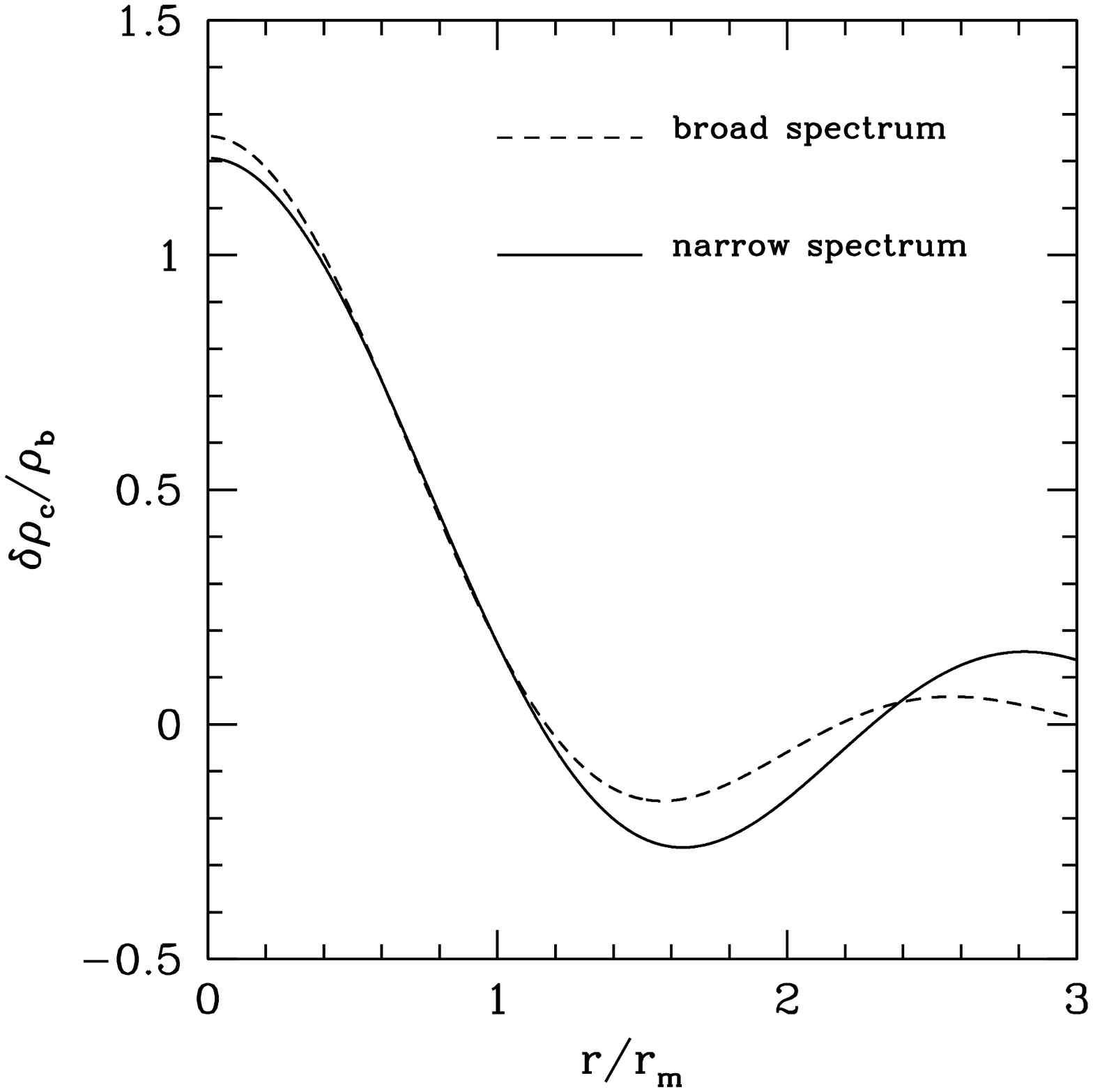}
\vspace{-2.5cm}
\caption{\label{fig1} This panel shows the critical density profile obtained from the narrow and broad power spectrum plotted against $r/r_m$. }
\end{figure*}

\subsection{Threshold of the primordial power spectrum}\label{Results}
We have so far discussed how to relate the abundance of PBHs to the primordial power spectrum in the case of rare peaks, $\nu_c\gg1$. 
We will see that generically \mbox{$\nu_c^2\propto {\cal P}^{-1}$}, and so the approximation of rare peaks, implying spherical symmetry, is 
intimately related to the linearity of the {\it mean} primordial perturbations. In the following, as benchmarks of power spectra generated during 
inflation, we will consider the case of a narrow power spectrum, and the opposite case of a broadspectrum, simplified as a top-hat distribution.

\subsubsection{Narrow power spectrum}
The first power spectrum which we consider is
\be
{\cal P}={\cal P}_0\ e^{-\frac{(k-k_p)^2}{2\sigma_{\cal P}^2}}\ ,
\ee 
in the limit of $k_p^2\gg\sigma_{\cal P}^2$. In this case one obtains the critical density profile plotted with a solid line in the left panel of 
figure \ref{fig1}. The parameters related to this profile are:  $k_*\simeq \sqrt{3} k_p$, $r_m\simeq \frac{2.7}{k_p}$ and 
$\tilde\sigma_0\simeq 0.7\sqrt{{\cal P}_0\sigma_{\cal P} k_p^3}$. Numerical simulations give  the following critical values: 
$\delta_c \simeq 0.51$, $\delta\rho_c/\rho_b\simeq1.2$,  ${\cal F}_c(0)\simeq 1.2/r_m^2\simeq 0.16 k_p^2$  which finally gives 
$\nu_c\simeq 0.22\sqrt{\frac{k_p}{\sigma_{\cal P}{\cal P}_0}}$.

To compare with previous literature and give an order of magnitude estimate, we can crudely approximate $\beta_f\sim e^{-\nu_c^2/2}$. 
For all of the dark matter being in PBHs of mass $10^{-16}\ M_\odot$, we would need \mbox{$\beta_f\sim 10^{-16}$} and therefore 
\mbox{${\cal P}_0\sim  7\times 10^{-4}\frac{k_p}{\sigma_{\cal P}}\gg 10^{-3}$} (this does not change significantly even up to 
\mbox{$M_{PBH}\sim 100\ M_\odot$}). Since for producing the seeds of PBHs from inflation one requires ${\cal P}_0\ll 1$, 
there is only a small margin for this kind of spectrum to work. 

Finally, using \eqref{Mpeak}, the PBHs formed by this spectrum are peaked at \mbox{$M_{PBH}\sim 0.8 M_H(t_m)$}.

\begin{figure*}[t!]
\centering
\vspace{-1.cm}
\includegraphics[scale=0.37]{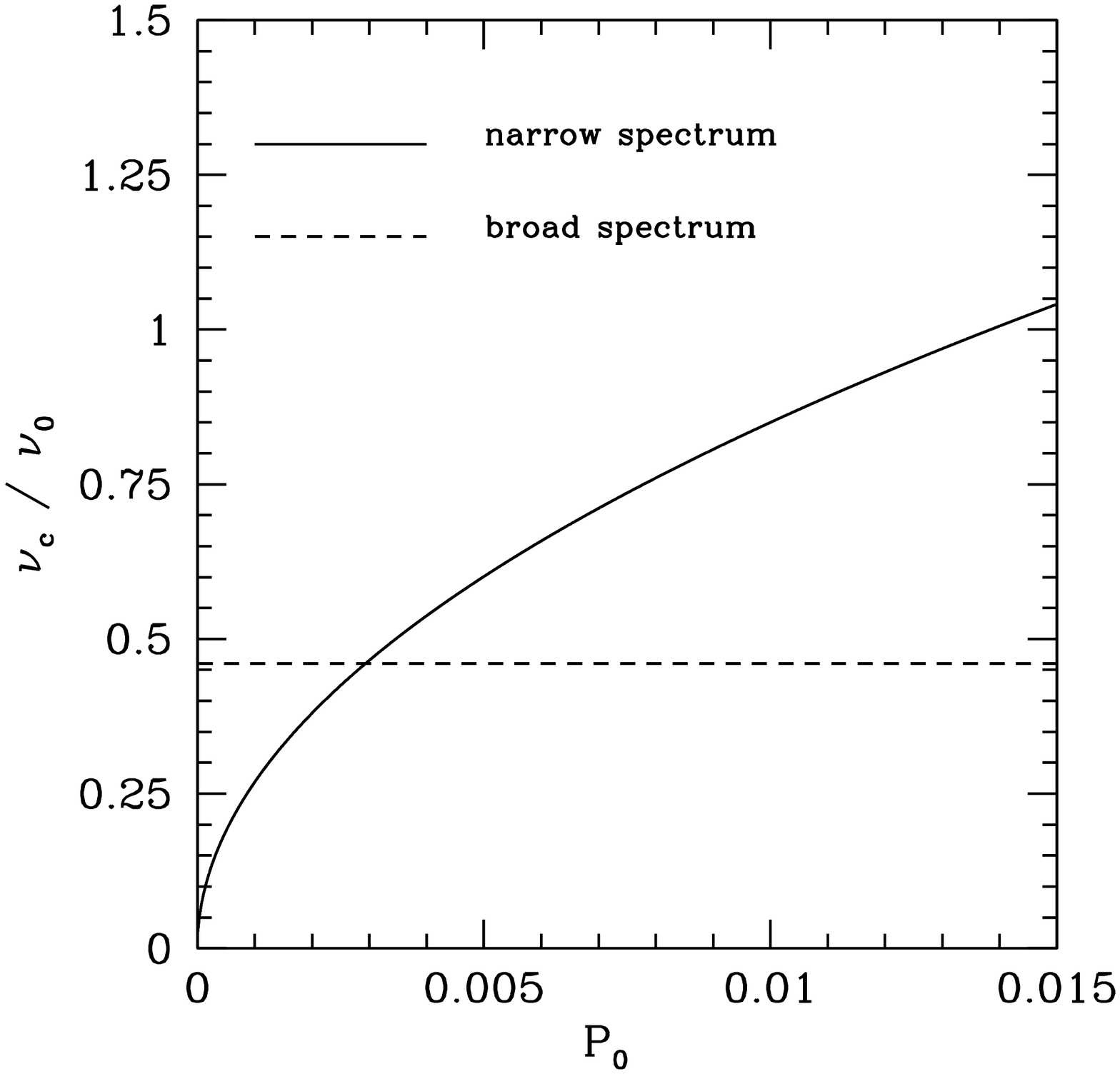}
\hspace{0.5cm}
\includegraphics[scale=0.37]{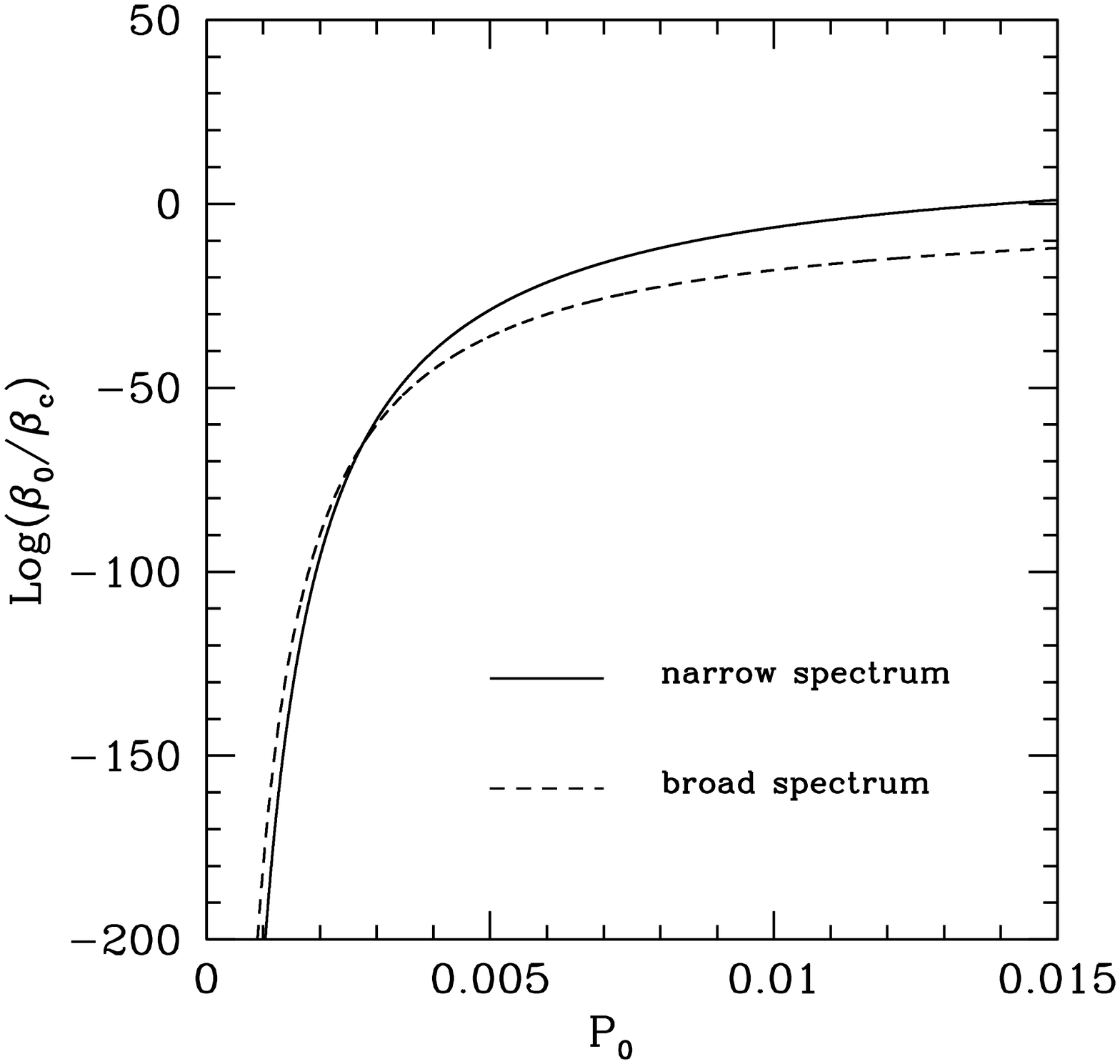}
\vspace{-2.5cm}
\caption{\label{fig2} The left panel shows the comparison between the threshold value $\nu_c$ and the value of $\nu_0$ 
calculated previously by the use of the Press-Schechter formalism, as a function of ${\cal P}_0$. The right panel shows instead the 
corresponding comparison between the approximated abundance of PBHs $\beta_c\equiv\beta(\nu_c)$ and $\beta_0\equiv\beta(\nu_0)$. 
For the broad spectrum the abundance is not fixed, while for the peaked one the ratio $k_p/\sigma_{\cal P}$ has been fixed by considering 
\mbox{$M_{PBH}\sim 10^{-16}\ M_{\odot}$}. 
}\end{figure*}

\subsubsection{Broad power spectrum} 
The second power spectrum considered is a top-hat with amplitude ${\cal P}_0$, extended between $[k_{min}$ and $k_{max}]$, with 
$k_{max}\gg k_{min}$.\footnote{Note that modes entering the cosmological horizon much later than the formation of the apparent horizon, 
which typically happens at the time $t_c\sim 10\, t_m$ \cite{muscoin}, will not participate in the black hole formation.}In this case, one obtains 
the critical density profile plotted with a dashed line in the left panel of figure \ref{fig1} with 
$k_*\simeq\sqrt{2}k_{max}$, \mbox{$r_m\simeq 3.5/k_{max}$} and \mbox{$\tilde\sigma_0\simeq 0.2\sqrt{{\cal P}_0}k_{max}^2$}.  

As one can see, the two profiles in units of $r/r_m$ are almost the same within a sphere of radius $r_m$. Therefore, numerical simulations give
basically the same values of $\delta_c$, $\delta\rho_c/\rho_b$ as in the previous case \cite{muscoin}. In terms of $k_{max}$ one then obtains 
${{\cal F}_c(0)}\simeq 0.10 k_{max}^2$ which finally gives $\nu_c\simeq 0.46 ({\cal P}_0)^{-1/2}$\,.
For $\beta_f\sim 10^{-16}$ we get ${\cal P}_0\sim 3\times 10^{-3}$, one order of magnitude smaller than the value $\sim 2\times  10^{-2}$
previously quoted in the literature, e.g. \cite{germani,ballesteros,cicoli}.  Therefore, for inflationary models generating this spectrum, 
it is found that PBHs are more likely to be produced than had been previously suggested in the literature. 

Finally, using \eqref{Mpeak}, the PBHs formed by this spectrum are peaked at a mass $M_{PBH}\sim 0.7 M_H(t_m)$.

\subsubsection{Discussion}
Although the expression for $\beta_f$ derived in eq. \eqref{int} differs in many aspects from the one used in the Press-Schechter approach, 
previously used in the literature (see for example \cite{carr2}), the main numerical difference comes from the discordant definitions of $\nu_c$: 
the PBH abundance was incorrectly related to the critical value of $\delta$ calculated at the edge of the over-density $r_0$ ignoring the profile 
dependence of the over-density. In particular,  $\delta_{0}\simeq 0.45$ corresponding to a Mexican-hat  profile \cite{muscop} was used earlier 
giving 
\be \label{nu_cold}
\nu_0 \simeq \frac{9}{4}\frac{\delta_0}{\sqrt{{\cal P}_0}}\simeq 1.01 {\cal P}_0^{-1/2} \,.
\ee
In figure \ref{fig1} we plot the two energy density profiles, corresponding to the narrow and broad peaks of the power spectrum, 
as a function of $r/r_m$. The two shapes are not significantly different because the profiles of the peak of the power spectrum for 
$k > k_p$ or $k > k_{max}$, which corresponds in real space to the region $r \lesssim r_m$, are very similar. Note that the Mexican-hat 
profile is very similar to the profiles drawn in Fig. \ref{fig1} and so the value of \mbox{$\delta_0\simeq 0.45$} is the relevant one for the profiles 
studied in this paper.

In the left frame of figure \ref{fig2} we plot the ratio between $\nu_c$ for the narrow and broad peak of the power spectrum calculated here 
with peak theory, and $\nu_0$ given by \eqref{nu_cold} as a function of ${\cal P}_0$, while in the right frame of figure \ref{fig2} we plot the 
ratio of the corresponding relative abundance $\beta_c/\beta_0$, with respect to the value of ${\cal P}_0$. For the narrow spectrum of both 
plots we have fixed the abundance \mbox{$\beta_f\sim 10^{-16}$} in the approximation $\beta_f(x)\sim e^{-x^2/2}$ because $\nu_c$ 
depends on $k_p/\sigma_{\cal P}$ and ${\cal P}_0$, while for the broad spectrum $\nu_c$ is only a function of ${\cal P}_0$ and the abundance 
is therefore varying with ${\cal P}_0$ along the dashed line. The intersection between the dashed line and the solid line gives the value of 
${\cal P}_0$ for the broad spectrum when $\beta_f\sim 10^{-16}$.

These plots shows clearly that the approach used previously was incorrect, with a value of $\nu_c \simeq 0.5\,\nu_0$ for the broad 
spectrum, while the difference for the narrow spectrum depends on ${\cal P}_0={\cal P}_0(\beta_f,k_p/\sigma_{\cal P})$. The error caused by 
using $\nu_0$ instead of $\nu_c$, once the variance of the spectrum is fixed,  becomes smaller for larger masses of PBHs because a larger 
value of $\beta_f$ corresponds to a larger value of ${\cal P}_0$. In particular, for the broad spectrum we have $\log(\beta_0/\beta_c)\simeq-2\nu_c$ 
which, for $M_{PBH}\sim10^{-16} M_\odot$, gives $\frac{\beta(\nu_0)}{\beta(\nu_c)}\sim 10^{-64}$ (see figure \ref{fig2}) while considering for 
example $M_{PBH}\sim100 M_\odot$, one has $\frac{\beta(\nu_0)}{\beta(\nu_c)}\sim 10^{-28}$.

Finally, let us stress that the cosmological horizon mass defined in our paper is defined at the horizon crossing time 
$a(t_m)H(t_m)r_m=1$. This mass generically differs from the one used in the literature which is calculated at the horizon crossing 
$a(t_k)H(t_k)/k=1$ of a characteristic mode $k$ (typically the oneassociated with the peak of the power spectrum). For example in the 
narrow spectrum $k=k_p$ and the mass calculated at $r_m$ is about $10$ times larger than the one calculated at $1/k_p$, as was also 
noted in \cite{Jaume}.

\section{Summary}
In the present letter we have re-analyzed the physics of PBH formation by combining peak theory \cite{peak} with the numerical 
analysis of gravitational collapse in the expanding universe \cite{muscoin}. We have computed the abundance of PBHs generated by a large 
peak in the primordial power spectrum of curvature perturbations. Characterizing the peak by its scale, amplitude and width, we heave shown 
that the abundance of PBHs is extremelydependent on the shape of the peak. The reason is that the threshold of the energy density peak for 
PBH formation depends strongly on the distribution of the real space over-density, which can be obtained, assuming Gaussian statistics, from 
the two-point correlation function of curvature perturbations. This crucial aspect had been overlooked in previous literature. 

Given the amplitude of the peak in the power spectrum at a particular scale, the abundance of PBHs generated by a narrow peak is exponentially 
smaller than the abundance generated by a broad one. In particular, to describe all of the dark matter with PBHs, using a top-hat profile of the peak 
in the power spectrum in Fourier space, the amplitude is an order of magnitude smaller than that previously calculated without taking into account 
the shape. Instead, for a narrow peak, as often assumed in the literature, one would need a much larger amplitude, which in many cases would 
invalidate the perturbative analysis of cosmological curvature perturbations. 

Our analysis has been done assuming negligible non-gaussianities in the initial conditions of the overdensity field. However, in certain cases, 
non-gaussianities of the curvature perturbations \cite{vicente} and/or non-gaussianities related to the non-linear relation between the curvature 
and over-density perturbations \cite{kehagias2} might give interesting contributions in the calculation of abundances of PBHs, both in terms of 
new statistics and for non-spherical deformations of the primordial perturbations. We leave these interesting questions for future research.

\vspace{0.1cm}

\begin{acknowledgments}
C.G. would like to thank Pierstefano Corasaniti and \mbox{Licia Verde} for discussions about peak theory and \mbox{Vicente Atal} 
for discussion about different inflationary models. IM would like to thank Misao Sasaki for discussions during the YITP long-term 
workshop ``Gravity and Cosmology 2018'', YITP-T-17-02. The Authors would like to thank Jaume Garriga, Juan Garcia-Bellido and 
John Miller for feedback and comments on the previous version of this paper. CG is supported by the Ramon y Cajal program and 
partially supported by the Unidad de Excelencia Mar\'ia de Maeztu Grant No. MDM-2014-0369 and by the national FPA2013-46570-C2-2-P 
and by the national FPA2016-76005-C2-2-P grants. IM is supported by the Unidad de Excelencia Mar\'ia de Maeztu Grant No. MDM-2014-0369.
\end{acknowledgments}


\end{document}